# In-plane *p*-wave coherence length in iron-based superconductors

E. F. Talantsev[1,2]

[1]M.N. Mikheev Institute of Metal Physics, Ural Branch, Russian Academy of Sciences, 18, S. Kovalevskoy St., Ekaterinburg, 620108, Russia

[2]NANOTECH Centre, Ural Federal University, 19 Mira St., Ekaterinburg, 620002, Russia

[*]E-mail: evgeny.talantsev@imp.uran.ru

**Abstract**

High-temperature superconductivity in iron-based layered compounds discovered by Hosono group (Kamihara *et al* 2006 *J. Am. Chem. Soc.* **128** 10012) is fascinating physical phenomenon which still has many unanswered questions. One of these questions is the superconducting gap symmetry in iron-based superconductors (IBS), for which the most agreed concept is multiple-band *s*-wave symmetry. Recently, an alternative concept of single-band *p*-wave symmetry has been proposed. To disprove/reaffirm the latter concept, in this paper we analyse temperature dependent in-plane coherence length, $\xi_{ab}(T)$, in FeSe, FeSe$_{1-x}$Te$_x$, Ba(Fe$_{1-x}$(Co/Ni)$_x$)$_2$As$_2$ and Ca$_{10}$(Pt$_4$As$_8$)((Fe$_{1-x}$Pt$_x$)$_2$As$_2$)$_5$ in order to extract the gap-to-critical-temperature ratio, $2\Delta(0)/k_B T_c$, and the specific-heat-jump ratio, $\Delta C/C$, in these compounds. In the result, we report that deduced ratios are in a good agreement with the concept of single-band *p*-wave superconductivity in these materials.

# In-plane *p*-wave coherence length in iron-based superconductors

## I. Introduction

The discovery of superconductivity in Fe-based compound of LaOFeP with superconducting transition near boiling point of liquid helium by Hosono's group [1] demonstrated that there is more deep link between the superconductivity and the ferromagnetism which both for decades considered as mutually exclusive phenomena. Further studies revealed more than dozen iron-based superconductors (IBS) families [2-5], including several systems which exhibit transition temperatures well above 21 K (i.e., the boiling point of hydrogen, which can be considered as a natural border from low- and high-temperature superconductors).

Superconducting gap symmetry in IBS is one of central question in understanding of the phenomenon in these compounds. Widely accepted view is that IBS exhibits multiple-band *s*-wave gap symmetry, in particular $s_\pm$-wave [6-12]. It should be noted, that this model has been proposed at very early stage of IBS studies, when precise experimental data for majority of IBS compounds has not been reported, and $s_\pm$-wave model is mainly originated from first-principles calculations, rather than from analysis of experimental data. However, recent thorough analysis of experimental temperature dependent superfluid density, $\rho_s(T)$, for several IBS systems [13] showed that single-gap *p*-wave model describes experimental data remarkably well, which is, in addition, required only 4 free-fitting parameters (i.e., transition temperature, $T_c$, ground state London penetration depth, $\lambda(0)$, ground state superconducting energy gap, $\Delta(0)$, and relative jump in specific heat at transition temperature, $\Delta C/C$), while $s_\pm$-wave model (or any other multiple-gaps model) requires more than twice free-fitting parameters (because, in addition to doubled number of mentioned above parameters, the model requires inter-band coupling constants). Due to a large number of free-fitting parameters, there is unlikely that all parameters are mutual independent, and, thus, an overfitting problem can be prominent when $s_\pm$-wave model is applied for the analysis of experimental data.

In this paper we attempt to reaffirm or disprove single-band *p*-wave superconductivity model in IBS compounds by the analysis of temperature dependent *c*-axis upper critical field, $B_{c2,c}(T)$ (i.e., when external applied magnetic field applied parallel to [001] direction of single crystal). This field is also called in-plane upper critical field and it is defined as [14]:

$$B_{c2,c}(T) = \frac{\phi_0}{2 \cdot \pi \cdot \xi_{ab}^2(T)} \tag{1}$$

where $\phi_0 = 2.068 \cdot 10^{-15}$ Wb is magnetic flux quantum, and $\xi_{ab}(T)$ is in-plane coherence length.

In the result, we conclude that all analysed IBS materials are single-band *p*-wave superconductors.

**II. Description of the approach**

To analyse $B_{c2,c}(T)$ data we use an approach which has been proposed in our previous works [15-17], and which is based on utilization a general relation between in-plane London penetration depth, $\lambda_{ab}(T)$, and in-plane the coherence length, $\xi_{ab}(T)$:

$$\kappa_c(T) = \frac{\lambda_{ab}(T)}{\xi_{ab}(T)} \tag{2}$$

where $\kappa_c(T)$ is in-plane Ginzburg-Landau parameter [14]. Thus, $B_{c2,c}(T)$ can be expressed in term of in-plane superfluid density, $\rho_{s,ab}(T)$:

$$\rho_{s,ab}(T) \equiv \frac{1}{\lambda_{ab}^2(T)} \tag{3}$$

$$B_{c2,c}(T) = \frac{\phi_0}{2 \cdot \pi \cdot \xi_{ab}^2(T)} = \frac{\phi_0 \cdot \kappa_c^2(T)}{2 \cdot \pi \cdot \lambda_{ab}^2(T)} = \frac{\phi_0}{2 \cdot \pi} \cdot \kappa_c^2(T) \cdot \rho_{s,ab}(T) \tag{4}$$

It should be stressed, that Eqs. 2-4 are valid when the order parameter phase fluctuations and the order parameter amplitude fluctuations are completely supressed and superconducting condensate is formed over the whole sample. This, in term of experimental measurements, means that Eqs. 2-4 are applicable when sample resistance under applied magnetic field, $B_{appl}$, is reached zero. In the literature, the upper critical field, $B_{c2}(T)$, defined at this strict criterion of $R(T,B) = 0\ \Omega$ is designated as the irreversibility field, $B_{irr}(T)$, while $B_{c2}(T)$ is defined at some value:

$$0 < \frac{R(T,B)}{R_{norm}} < 1 \qquad (5)$$

where $R_{norm}$ is normal state resistance [18-22]. Despite of this, hereafter, we will use the designation of $B_{c2}(T)$ for the field at which initial stage of dissipation has been registered in experiment. It should be noted that not for all IBS materials and not all measurements (and, especially, for measurements at high pulsed magnetic field [23]) the criterion of $R = 0\ \Omega$ can be used because of experimental challenges, in these cases the lowest possible value of $\frac{R(T,B)}{R_{norm}(T)} \lesssim 0.1$ will be applied to define $B_{c2}(T)$ from $R(T,B)$ curves as described below.

Thus, our approach is based on the assumption that the superfluid density, $\rho_s(T) \equiv \frac{1}{\lambda^2(T)}$, for which the most accurate, precise and widely accredited measuring technique is the muon-spin rotation spectroscopy (μSR) [24-28], can be converted into $B_{c2}(T) \sim \frac{1}{\xi^2(T)}$ by employing multiplicative pre-factor $\frac{\phi_0}{2\cdot\pi} \cdot \kappa_c^2(T)$ (Eq. 4).

Due to $\rho_s(T) \equiv \frac{1}{\lambda^2(T)}$ can be very precisely calculated for different types of superconducting gap symmetries [29-33], the accuracy of the approach is based on general form of temperature dependent Ginzburg-Landau parameter κ(T), for which we used [15-17] an expression proposed by Gok'kov [34,35]:

$$\kappa\left(\frac{T}{T_c}\right) = \kappa(0) \cdot \left(1 - 0.243 \cdot \left(\frac{T}{T_c}\right)^2 + 0.039 \cdot \left(\frac{T}{T_c}\right)^4\right) \qquad (6)$$

It can be noted that alternative temperature dependences for $\kappa\left(\frac{T}{T_c}\right)$ were proposed, from which we can mentioned:

$$\kappa\left(\frac{T}{T_c}\right) = \kappa(0) \cdot \left(1 - 0.31 \cdot \left(\frac{T}{T_c}\right)^2 \cdot \left(1 - 1.77 \cdot \ln\left(\frac{T}{T_c}\right)\right)\right) \qquad (7)$$

proposed by Summers *et al* [36], and:

$$\kappa\left(\frac{T}{T_c}\right) = \kappa(0) \cdot \left(\frac{1-\left(\frac{T}{T_c}\right)^{1.52}}{1-\left(\frac{T}{T_c}\right)^2}\right) \qquad (8)$$

proposed by Godeke *et al* [37]. However, Eq. 7 has infinite term $ln\left(\frac{T}{T_c}\right)$ at $T \to 0\ K$, and Eq. 8 has uncertainty of $\frac{0}{0}$ at $T \to T_c$, which both are very unlikely features for reasonably smooth and simple physical parameter κ(T). Moreover, despite differences in formulations (Eqs. 6-8), all these temperature dependences give similar $\frac{\kappa(T)}{\kappa(0)}$ values for whole temperature range (Fig. 1).

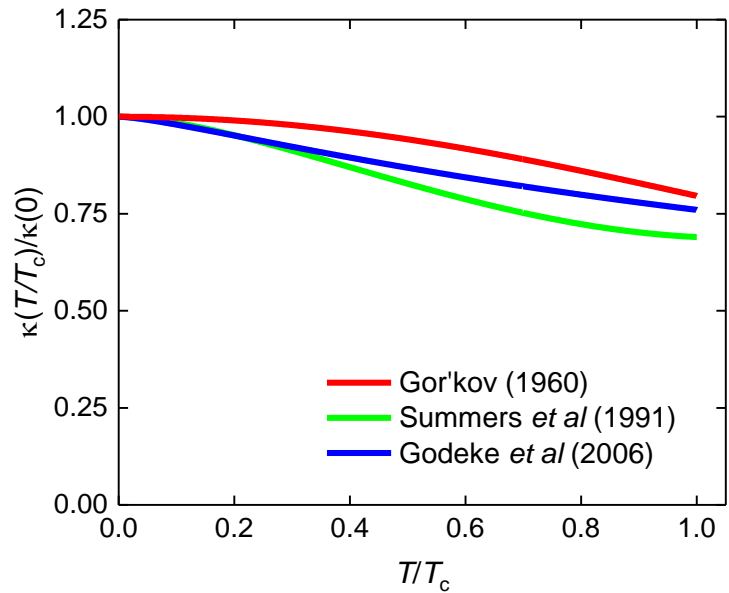

**Figure 1.** Temperature dependence of Ginzburg-Landau parameter $\frac{\kappa\left(\frac{T}{T_c}\right)}{\kappa(0)}$ proposed by Gor'kov [34,35], Summers *et al* [36], and Godeke *et al* [37].

To present more evidences that Eq. 4 is valuable approach to study the upper critical field, $B_{c2}(T)$, in Fig. 2 we have undertaken a manual scaling of $\frac{B_{c2,c}(T)}{B_{c2,c}(0)}$ data for single crystal La$_{1.83}$Sr$_{0.17}$CuO$_4$ reported by Ando *et al* [38] in their Fig. 4(b) (for which we used $\frac{R(T,B)}{R_{norm}(T)} =$ 0.1, $T_c$ = 35 K, and $B_{c2,c}(0)$ = 35 T) and $\frac{\kappa_c^2\left(\frac{T}{T_c}\right)\cdot\rho_{s,ab}\left(\frac{T}{T_c}\right)}{\kappa_c^2(0)\cdot\rho_{s,ab}(0)}$ for which we used μSR data reported by Wojek et al [28] for $La_{1.83}Sr_{0.17}CuO_4$ single crystal. It can be seen (Fig. 2) that there is a very good agreement between manually scaled $\frac{B_{c2,c}\left(\frac{T}{T_c}\right)}{B_{c2,c}(0)}$ and $\frac{\kappa_c^2(T)\cdot\rho_{s,ab}\left(\frac{T}{T_c}\right)}{\kappa_c^2(0)\cdot\rho_{s,ab}(0)}$ data for all three $\kappa\left(\frac{T}{T_c}\right)$ (Eqs. 6-8).

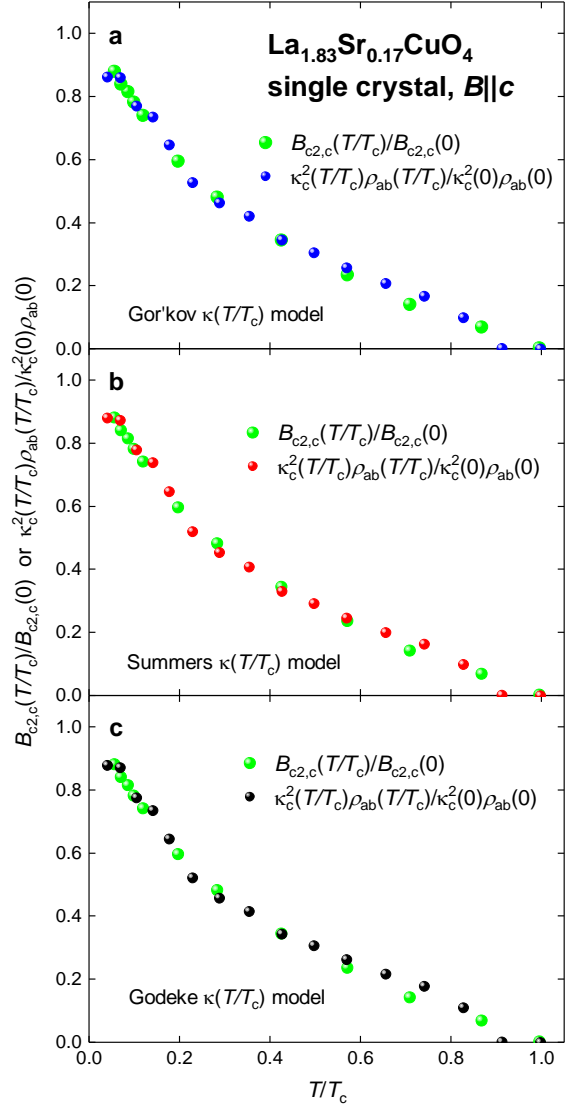

**Figure 2.** $\frac{B_{c2,c}\left(\frac{T}{T_c}\right)}{B_{c2,c}(0)}$ and $\frac{\kappa_c^2\left(\frac{T}{T_c}\right)\cdot\rho_{s,ab}\left(\frac{T}{T_c}\right)}{\kappa_c^2(0)\cdot\rho_{s,ab}(0)}$ data for single crystal $La_{1.83}Sr_{0.17}CuO_4$. External magnetic field applies parallel to $c$-axis of the crystal. Raw $B_{c2,c}(T)$ data is from Ref. 38, raw $\rho_{s,ab}(T)$ data is from Ref. 28. $\rho_{s,ab}\left(\frac{T}{T_c}\right)$ is recalculated in $B_{c2,c}(\frac{T}{T_c})$ by temperature dependent Ginzburg-Landau parameter $\kappa_c\left(\frac{T}{T_c}\right)$ proposed by (a) Gor'kov [34,35], (b) Summers et al [36], and (c) Godeke et al [37].

In our previous work [13], we showed that the self-field critical current density, $J_c(sf,T)$, in thin (when the film thickness, $2b$, is less than $c$-axis London penetration depth, $\lambda_c(0)$) $c$-axis oriented films of IBS described by:

$$J_c(sf,T) = \frac{\phi_0}{4\pi\mu_0} \cdot (ln(\kappa_c) + 0.5) \cdot \rho_{ab,p,\perp}^{1.5}(T). \tag{9}$$

where $\rho_{ab,p,\perp}(T) \equiv \frac{1}{\lambda_{ab,p,\perp}^2(T)}$ is polar $\mathbf{A}\perp l$ $p$-wave superfluid density in $ab$-plane, where $\mathbf{A}$ is vector potential and $l$ is the gap axis orientation. Based on this result, there is a logical expectation that the coherence length in ab-plane should be also polar $\mathbf{A}\perp l$ $p$-wave, $\xi_{ab,p,\perp}(T)$. This is what we report in this paper.

Thus, main result of this paper is demonstrated in Figure 3, where in panel (a) we show reduced $\frac{\rho_s\left(\frac{T}{T_c}\right)}{\rho_s(0)} \equiv \frac{\lambda^2(0)}{\lambda^2\left(\frac{T}{T_c}\right)}$ curves for single-band weak-coupling $s$-, $d$- and polar $\mathbf{A}\perp l$ $p$-wave gap symmetries. In Fig. 3(b), these three curves are multiplied by $\kappa^2\left(\frac{T}{T_c}\right)$ factor proposed by Gor'kov (Eq. 6), which give weak-coupling curves for $\frac{B_{c2}\left(\frac{T}{T_c}\right)}{B_{c2}(0)}$. In Fig. 3(b) we also show manual scaling of $B_{c2,c}(T)$ data ($B_{appl}$ is oriented along [001] direction) for V$_3$S single crystal reported by Foner and McNiff [39], and $B_{c2,c}(T)$ for Ba(Fe$_{1-x}$Co$_x$)$_2$As$_2$ (x = 0.08) epitaxial thin film deposited on (001) Fe/MgO single crystal reported by Hänisch *et al* [40] in their Fig. 8(a).

It can be seen (Fig. 3,b) that $B_{c2,c}(T)$ data for $s$-wave superconductor V$_3$Si is only slightly deviate from the calculated weak-coupling $\frac{\kappa_c^2\left(\frac{T}{T_c}\right)\cdot\rho_{s,ab}\left(\frac{T}{T_c}\right)}{\kappa_c^2(0)\cdot\rho_{s,ab}(0)}$ curve, due to V$_3$Si is moderately strong coupling superconductor (with the gap-to-critical-temperature ratio of $\frac{2\cdot\Delta(0)}{k_B\cdot T_c} = 3.8$ [41]), which is slightly larger than weak-coupling value of $\frac{2\cdot\Delta(0)}{k_B\cdot T_c} = 3.53$ for which the curve was calculated. In section 3.1 we perform numerical fit of this dataset (rather than manual scaling) which confirms moderately-strong coupling electron-phonon interaction in V$_3$Si superconductor.

It should be also stressed that manual scaling (Fig. 3,b) $B_{c2,c}(T)$ data for Ba(Fe$_{1-x}$Co$_x$)$_2$As$_2$ [39] is nicely match weak-coupling $\frac{\kappa_c^2\left(\frac{T}{T_c}\right)\cdot\rho_{s,ab}\left(\frac{T}{T_c}\right)}{\kappa_c^2(0)\cdot\rho_{s,ab}(0)}$ curve for polar $\mathbf{A}\perp\mathbf{l}$ $p$-wave for which the gap-to-critical-temperature ratio is $\frac{2\cdot\Delta(0)}{k_B\cdot T_c} = 4.92$. We also perform numerical fit of this $B_{c2,c}(T)$ dataset in Section 3.2.

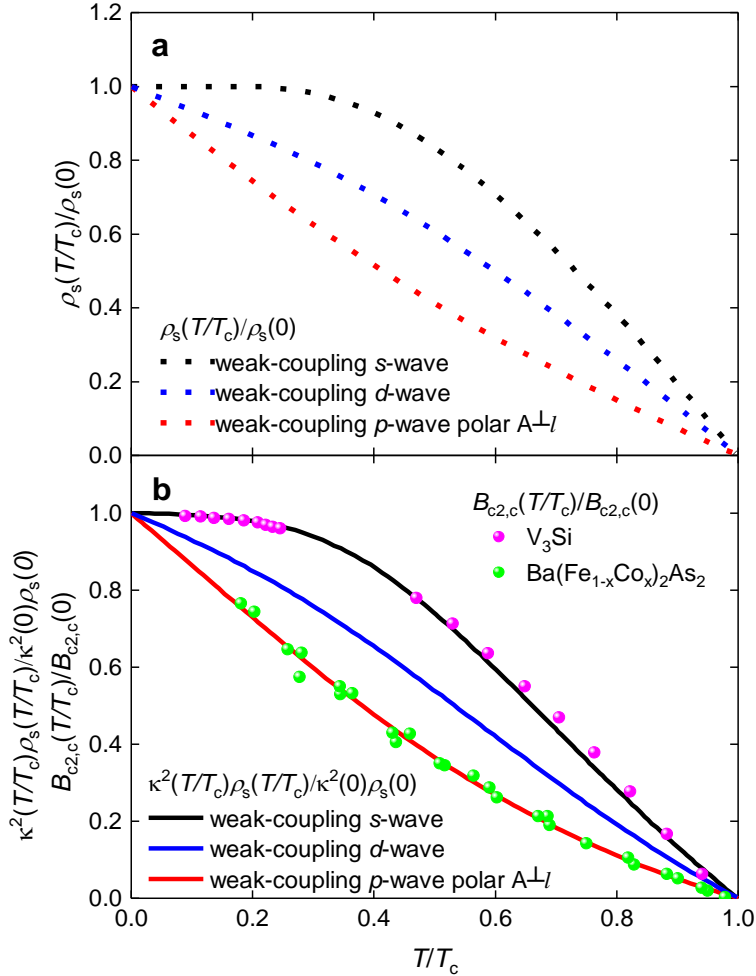

**Figure 3.** Temperature dependences for (a) reduced superfluid density $\frac{\rho_s\left(\frac{T}{T_c}\right)}{\rho_s(0)}$ and (b) $\frac{\kappa_c^2\left(\frac{T}{T_c}\right)\cdot\rho_{s,ab}\left(\frac{T}{T_c}\right)}{\kappa_c^2(0)\cdot\rho_{s,ab}(0)}$ for single-band weak-coupling $s$-, $d$-, and polar $\mathbf{A}\perp\mathbf{l}$ $p$-wave gap symmetries. Manual scaling of experimental $B_{c2,c}(T)$ data for V$_3$Si single crystal [39] (scaling parameters are: $T_c$ = 17 K, $B_{c2,c}(0)$ = 18.8 T) and for Ba(Fe$_{1-x}$Co$_x$)$_2$As$_2$ (x = 0.08) [40] (scaling paramaters are: $T_c$ = 23.1 K, $B_{c2,c}(0)$ = 47.5 T) are shown in panel (b).

While our primarily finding is shown in Fig. 3, in the next section we deduce exact values for the ground state coherence length, $\xi(0)$, ground state energy gap, $\Delta(0)$, and the relative jump in specific heat, $\Delta C/C$, at superconducting transition temperature, and transition temperature, $T_c$, for V$_3$Si single crystal by fitting experimental $B_{c2,c}(T)$ data [39] to the equation [14]:

$$B_{c2,c}(T) = \frac{\phi_0}{2\cdot\pi\cdot\xi_{ab}^2(0)} \cdot \left(\frac{1.77 - 0.43\cdot\left(\frac{T}{T_c}\right)^2 + 0.07\cdot\left(\frac{T}{T_c}\right)^4}{1.77}\right)^2 \cdot \left[1 - \frac{1}{2\cdot k_B\cdot T}\cdot \int_0^\infty \frac{d\varepsilon}{\cosh^2\left(\frac{\sqrt{\varepsilon^2 + \Delta^2(T)}}{2\cdot k_B\cdot T}\right)}\right] \quad (9)$$

where $k_B$ is the Boltzmann constant, and the amplitude of temperature dependent superconducting gap, $\Delta(T)$, is given by [29,30]:

$$\Delta(T) = \Delta(0)\cdot \tanh\left[\frac{\pi\cdot k_B\cdot T_c}{\Delta(0)}\cdot\sqrt{\eta\cdot\frac{\Delta C}{C}\cdot\left(\frac{T_c}{T} - 1\right)}\right] \quad (10)$$

where $\Delta C/C$ is the relative jump in electronic specific heat at $T_c$, and $\eta = 2/3$ for *s*-wave superconductors.

In following sections, where $B_{c2,c}(T)$ data of IBS is analyzed, we use fitting equation:

$$B_{c2,c}(T) = \frac{\phi_0}{2\cdot\pi\cdot\xi_{ab}^2(0)} \cdot \left(\frac{1.77 - 0.43\cdot\left(\frac{T}{T_c}\right)^2 + 0.07\cdot\left(\frac{T}{T_c}\right)^4}{1.77}\right)^2 \cdot \left[1 - \frac{3}{4\cdot k_B\cdot T}\cdot \int_0^1 w_\perp(x)\cdot\left(\int_0^\infty \frac{d\varepsilon}{\cosh^2\left(\frac{\sqrt{\varepsilon^2 + \Delta_p^2(T)\cdot f_p^2(x)}}{2\cdot k_B\cdot T}\right)}\right)\cdot dx\right] \quad (11)$$

where subscripts *p* and ⊥ designate polar perpendicular case of *p*-wave symmetry for which general gap function is given by [29,30]:

$$\Delta(\hat{\boldsymbol{k}}, T) = \Delta(T) f(\hat{\boldsymbol{k}}, \hat{\boldsymbol{l}}) \quad (12)$$

where, $\Delta(T)$ is the superconducting gap amplitude, $\boldsymbol{k}$ is the wave vector, and $\boldsymbol{l}$ is the gap axis. The function of $w_\perp(x)$ in Eq. 11 is:

$$w_\perp(x) = (1-x^2)/2 \quad (13)$$

The gap amplitude in Eq. 12 is given by Eq. 10, but $\eta$ in Eq. 10 is given by [29,30]:

$$\eta_p = \frac{2}{3}\cdot \frac{1}{\int_0^1 f_p^2(x)\cdot dx} \quad (14)$$

where

$$f_p(x) = x \qquad (15)$$

More details about geometries can be found elsewhere [13,29,30]. By substituting Eqs. 10, 13-15 in Eq. 11, one can fit experimental $J_c(sf,T)$ data to polar $\mathbf{A}\perp\mathbf{l}$ $p$-wave gap symmetries to deduce $\xi_{ab}(0)$, $\Delta(0)$, $\Delta C/C$, $T_c$ as free-fitting parameters and $\frac{2\Delta(0)}{k_B T_c}$ can be calculated.

### III. Results and Discussion

#### 3.1. V$_3$Si single crystal

Foner and McNiff reported $B_{c2,c}(T)$ data for high-quality V$_3$Si single crystal (which was also studied by Maita and Bucher [42]) in their Fig. 1 [39]. Orlando *et al* [41] analysed this $B_{c2,c}(T)$ dataset by Werthamer-Helfand-Hohenberg (WHH) theory [42,43], and we use Eqs. 9,10 to fit the same dataset. The fit result is shown in Fig. 4.

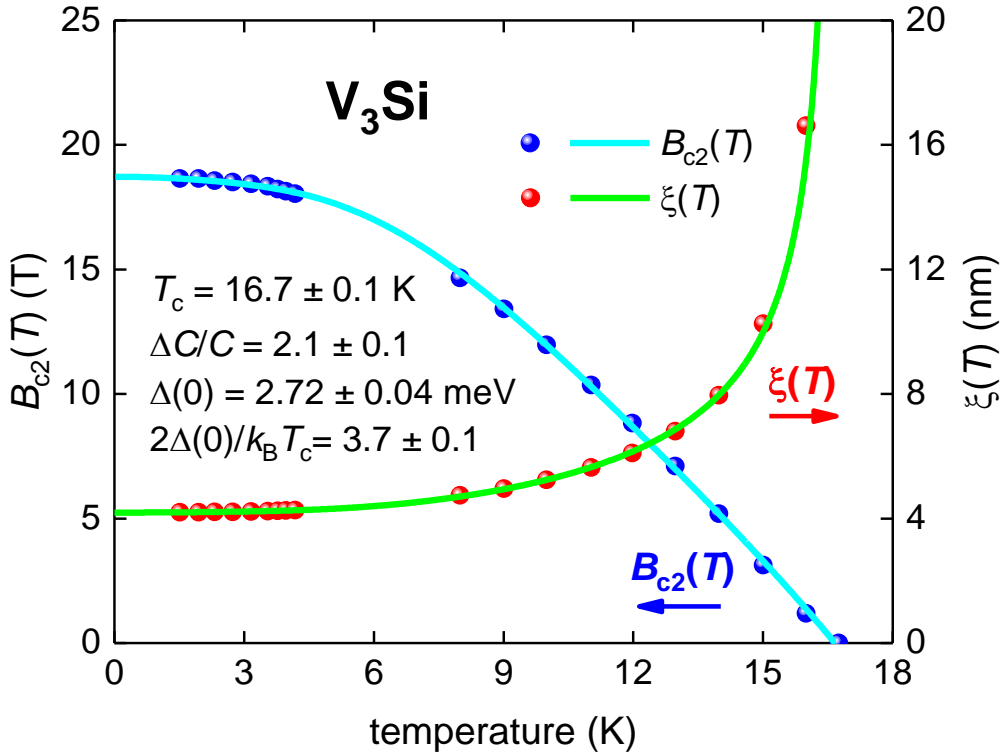

**Figure 4.** Upper critical field, $B_{c2}(T)$ (raw data from Ref. 39), and fit to Eqs. 9,10 for V$_3$Si single crustal. Raw $B_{c2}(T)$ data is from Ref. 37. Deduced $\xi(0) = 4.19 \pm 0.01$ nm. Fit quality is $R = 0.9998$.

It is important to note that deduced the gap-to-critical-temperature ratio of $\frac{2\cdot\Delta(0)}{k_B\cdot T_c} = 3.7 \pm 0.1$ within uncertainty range is equal to the value of $\frac{2\cdot\Delta(0)}{k_B\cdot T_c} = 3.8$ reported by Orlando *et al* [41] for the same crystal. This is an evidence which supports the use of basic equation Eq. 4.

### 3.2. Ba(Fe$_{1-x}$Co$_x$)$_2$As$_2$ (x = 0.08) thin film

Polar $\mathbf{A}\perp l$ p-wave gap pairing symmetry has weak-coupling limits of $\frac{2\cdot\Delta(0)}{k_B\cdot T_c} = 4.92$ and $\frac{\Delta C}{C} = 0.79$. A fit to Eqs. 10,11,13-15 raw experimental $B_{c2,c}(T)$ data reported by Hänisch *et al* [40] in their Fig. 8(a) (which has been designated as $H_{irr}(T)$ [40]) for high-quality epitaxial c-axis oriented film of Ba(Fe$_{1-x}$Co$_x$)$_2$As$_2$ (x = 0.08) is shown in Fig. 5.

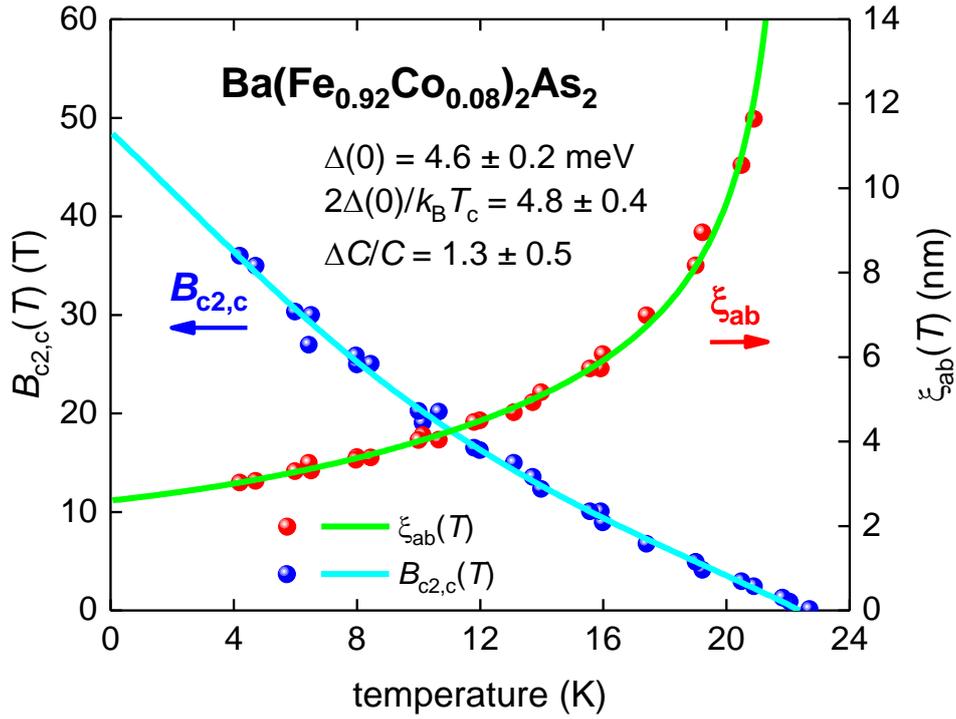

**Figure 5.** *c*-axis upper critical field, $B_{c2,c}(T)$, and data fit to Eqs. 10,11,13-15 for epitaxial c-axis oriented film of Ba(Fe$_{1-x}$Co$_x$)$_2$As$_2$ (x = 0.08) reported by Hänisch *et al* [40]. Deduced $T_c$ = 22.4 ± 1.0 K and $\xi_{ab}(0)$ = 2.61 ± 0.05 nm. Fit quality is $R = 0.9506$.

Deduced $\frac{2\cdot\Delta(0)}{k_B\cdot T_c} = 4.8 \pm 0.2$ and $\frac{\Delta C}{C} = 1.3 \pm 0.5$ are within uncertainty ranges are in an agreement with weak-coupling scenario in Ba(Fe$_{1-x}$Co$_x$)$_2$As$_2$ (x = 0.08).

### 3.3. Ba(Fe$_{1-x}$Co$_x$)$_2$As$_2$ (x = 0.16) single crystal

Kano *et al* [45] reported $B_{c2,c}(T)$ data for Ba(Fe$_{1-x}$Co$_x$)$_2$As$_2$ (x = 0.16) single crystal in their Fig. 5. Despite the authors used $\frac{R(T,B)}{R_{norm}(T)} = 0.5$ criterion for $B_{c2,c}(T)$ definition, we use their data because experimental $R(T,B)$ curves show sharp transition (Figs. 2,3 [45]) and at very high pulsed magnetic field which were employed in the study (see Fig. 2 [45]), it is difficult to employ more strict criterion. The fit shows that twice larger Co-doping does not make an effect on absolute value of the coherence length, $\xi_{ab}(0)$, in comparison with Ba(Fe$_{1-x}$Co$_x$)$_2$As$_2$ (x = 0.08) (Section 3.2), but it strengths the pairing interaction to moderately strong level, which is manifested by $\frac{2\cdot\Delta(0)}{k_B\cdot T_c} = 5.7 \pm 0.1$ and $\frac{\Delta C}{C} = 1.8 \pm 0.2$ (Fig. 6).

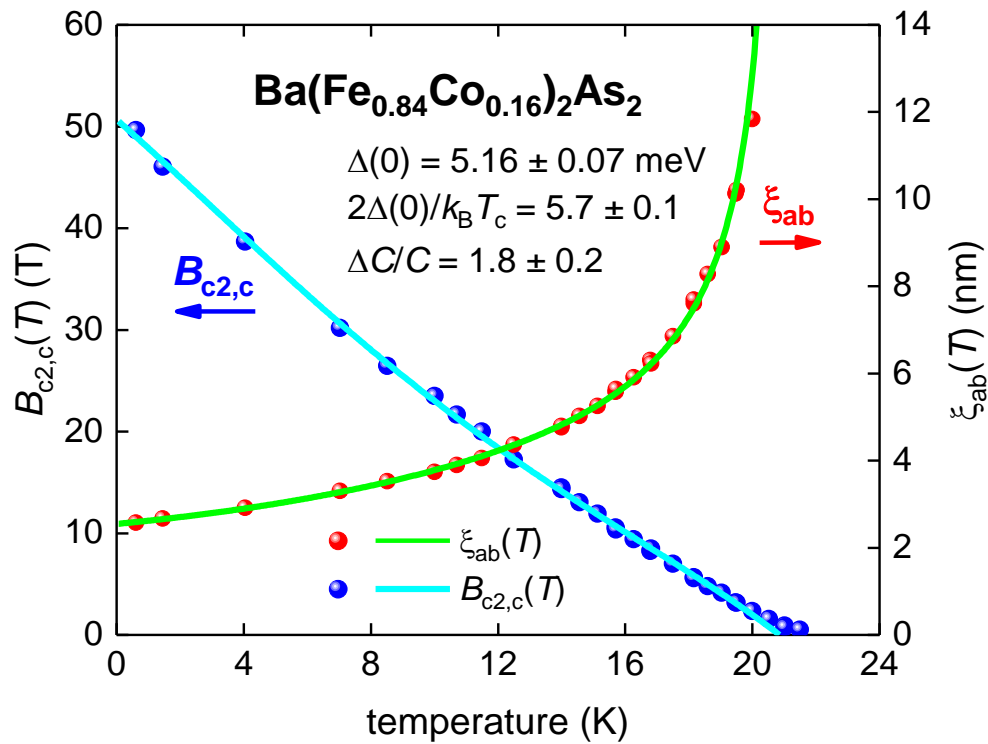

**Figure 6.** *c*-axis upper critical field, $B_{c2,c}(T)$ (raw data reported by Kano *et al* [45]) and data fit to Eqs. 10,11,13-15 for Ba(Fe$_{1-x}$Co$_x$)$_2$As$_2$ (x = 0.16) single crystal. Deduced $T_c = 20.9 \pm 0.2$ K and $\xi_{ab}(0) = 2.55 \pm 0.03$ nm. Fit quality is $R = 0.9926$.

### 3.4. Ba(Fe$_{1-x}$Ni$_x$)$_2$As$_2$ (x = 0.046) single crystal

Wang *et al* [46] studied the upper critical field in series of Ba(Fe$_{1-x}$Ni$_x$)$_2$As$_2$ (x = 0.0325-0.11) single crystals and fit $B_{c2}(T)$ data to two-bands WHH model. Wang *et al* [46] in their Fig. 2,b show raw $R(T,B)$ curve from which by applying $\frac{R(T,B)}{R_{norm}(T)} = 0.02$ criterion we deduced $B_{c2,c}(T)$ data which is shown in Fig. 7. Due to limited number of experimental $B_{c2,c}(T)$ data points we fix one of parameters, $T_c$, to its experimental value of 18.6 K and fit data to Eqs. 10,11,13-15. The fit reveals moderate level of coupling strength in this compound.

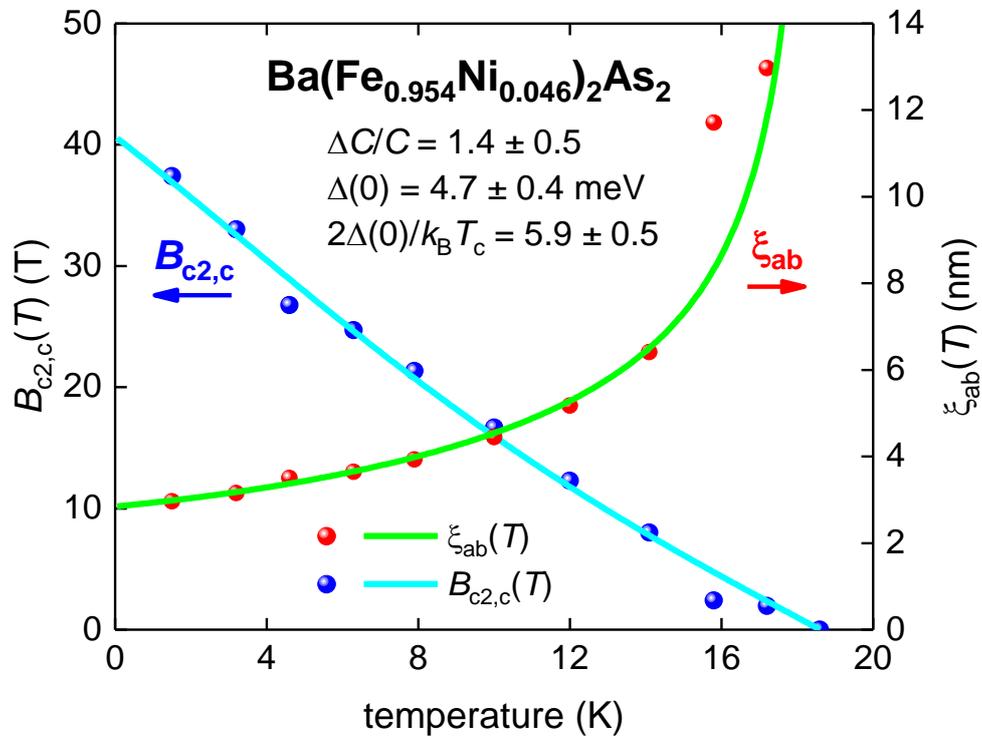

**Figure 7.** *c*-axis upper critical field, $B_{c2,c}(T)$ (raw data reported by Wang *et al* [46]) and data fit to Eqs. 10,11,13-15 for Ba(Fe$_{1-x}$Ni$_x$)$_2$As$_2$ (x = 0.046) single crystal. $T_c$ = 18.6 K (fixed) and $\xi_{ab}(0) = 2.97 \pm 0.05$ nm. Fit quality is $R$ = 0.8493.

### 3.5. FeSe single crystal

FeSe is iron-based superconductor which has the simplest crystalline structure (which whole IBS family) and simultaneously exhibiting the most intriguing property to be thinning down to atomic thickness, its critical temperature shows unprecedented rise in $T_c$ [47-50]. In this paper we analyse $B_{c2,c}(T)$ data for bulk FeSe crystals which exhibit $T_c \sim$ 8-9 K.

In Fig. 8 we show $B_{c2,c}(T)$ data reported by Vedeneev *et al* [51] (defined by zero resistance criterion of $R(T,B) = 0.0\ \Omega$, see Fig. 6(a) [51]) and fit this dataset to Eqs. 10,11,13-15. Due to $B_{c2,c}(T)$ data does not have reasonable number of data points at high reduced temperature, for fit we fixed $T_c$ to observed value of 8.65 K.

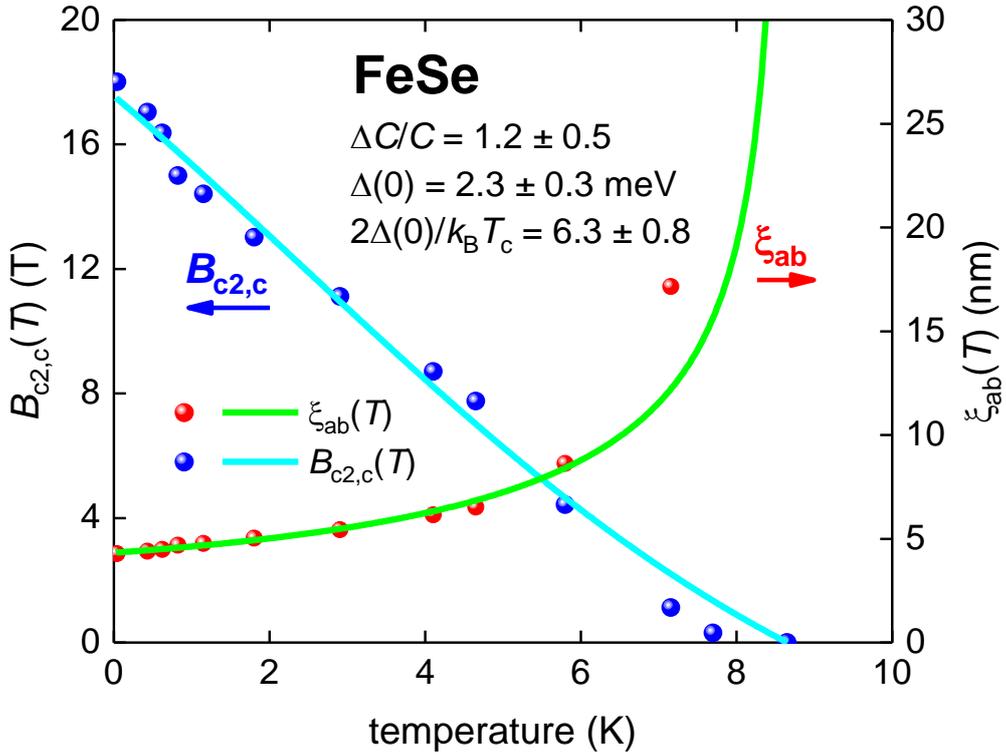

**Figure 8.** *c*-axis upper critical field, $B_{c2,c}(T)$ (raw data reported by Vedeneev *et al* [51]) and data fit to Eqs. 10,11,13-15 for FeSe single crystal. $T_c$ = 8.65 K (fixed) and $\xi_{ab}(0)$ = 4.34 ± 0.05 nm. Fit quality is $R$ = 0.8817.

The fit reveals reasonably strong electron-phonon coupling which characterized by the ratio of $\frac{2 \cdot \Delta(0)}{k_B \cdot T_c} = 6.3 \pm 0.8$, which corresponds to the amplitude of the superconducting energy gap of $\Delta(0) = 2.3 \pm 0.3$ meV, which is in a good agreement with independent measurement the amplitude of the superconducting gap in FeSe [52]. The ratio of $\Delta C/C$ is within the uncertainty is in a good agreement with independent measurements of 1.55 [53].

Similar result (Fig. 9) is obtained for the upper critical field for single crystal FeSe reported by Sun et al [54] (which designates as $H_{irr}$ in Fig. 5 [54]). For the fit we fixed $T_c = 9$ K to its experimental value (Fig. 1 [54]).

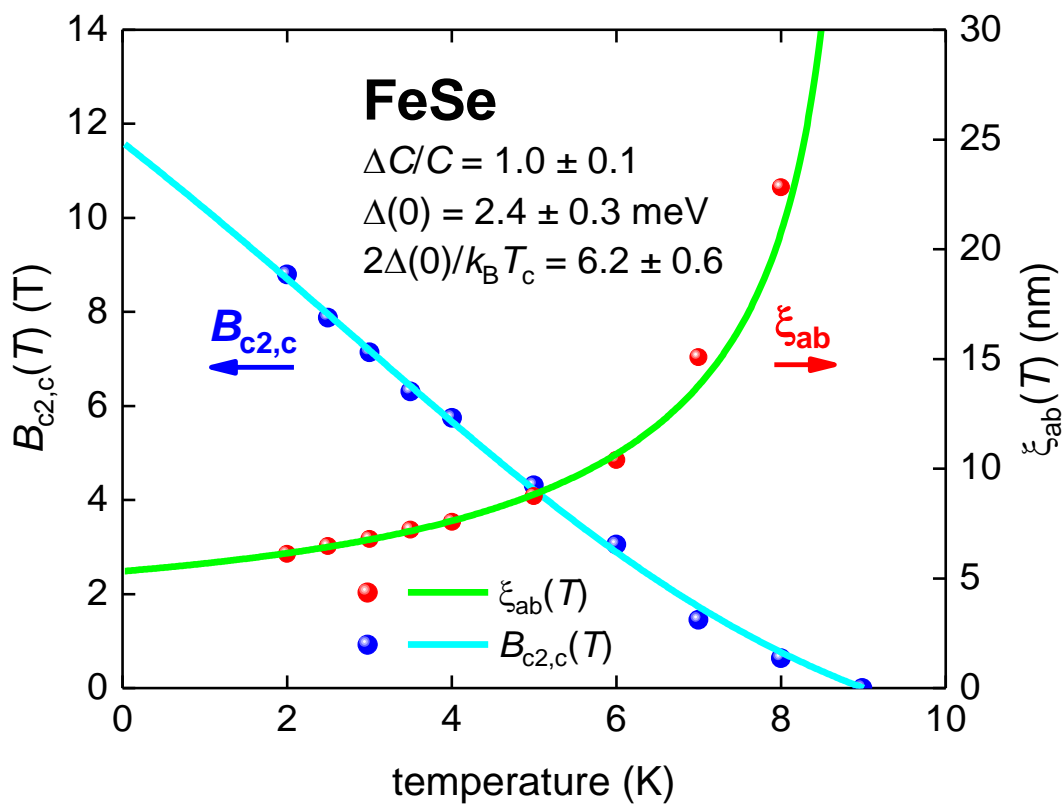

**Figure 9.** *c*-axis upper critical field, $B_{c2,c}(T)$ (raw data reported by Sun *et al* [54]) and data fit to Eqs. 11,13-15 for FeSe single crystal. $T_c = 9.0$ K (fixed) and $\xi_{ab}(0) = 5.35 \pm 0.06$ nm. Fit quality is $R = 0.9624$.

### 3.6. Fe$_{1+y}$Te$_{1-x}$Se$_x$ single crystal (x = 0.39, y = 0.02)

Lei *et al* [55] reported one of the first study the upper critical field, $B_{c2}(T)$, in tellurium doped FeSe single crystals, Fe$_{1+y}$Te$_{1-x}$Se$_x$. Extended review for these studies reported by Sun *et al* [56]. The fit to for which in Fig. 10 we show $B_{c2}(T)$ data and fit to Eqs. 10,11,13-15 for single crystal $Fe_{1+y}Te_{1-x}Se_x$ single crystal (x = 0.39, y = 0.02) reported by Lei *et al* [55] for which we employed the criterion of $\frac{R(T,B)}{R_{norm}(T)} = 0.02$ to deduce $B_{c2}(T)$ data from $R(T,B)$ curves reported by Lei *et al* [55].

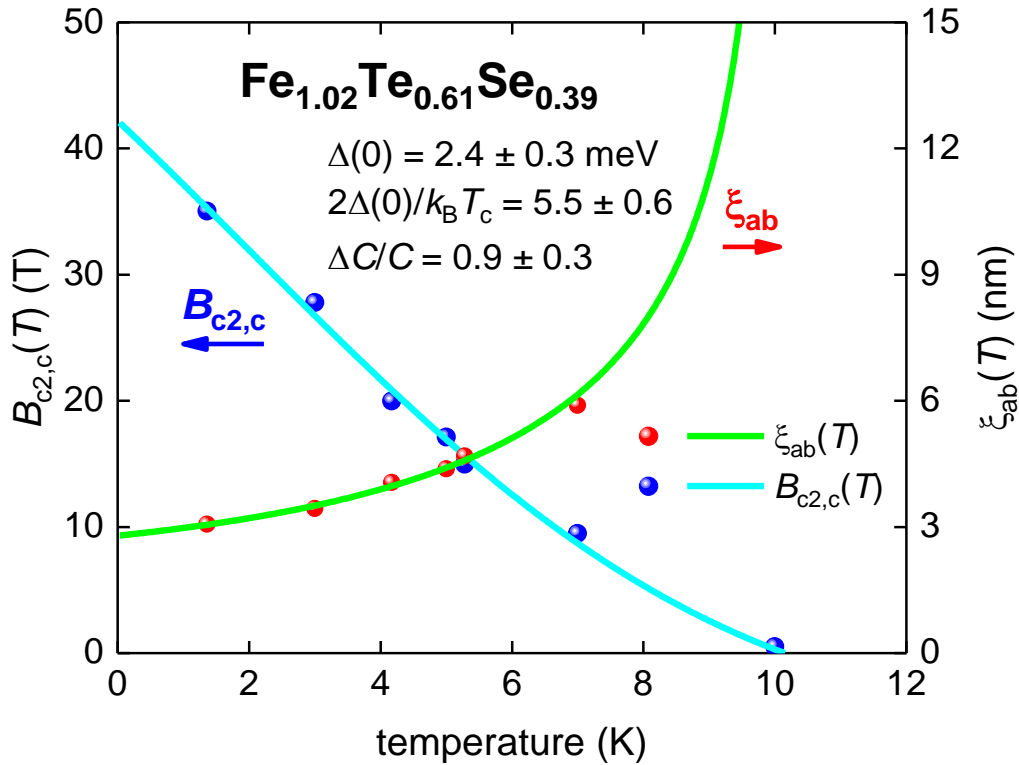

**Figure 10.** *c*-axis upper critical field, $B_{c2,c}(T)$ (raw data reported by Lei *et al* [55]) and data fit to Eqs. 11,13-15 for $Fe_{1+y}Te_{1-x}Se_x$ single crystal (x = 0.39, y = 0.02) single crystal. $T_c$ = 10.1 K (fixed) and $\xi_{ab}(0) = 5.35 \pm 0.06$ nm. Fit quality is $R = 0.8578$.

To perform fit we fixed $T_c$ to its experimental value of 10.1 K. It can be seen in Fig. 10 that deduced parameters are in good agreement with weak-coupling limit of *p*-wave gap symmetry. It should be also mentioned that this first reported $B_{c2}(T)$ data already showed that neither *s*-, nor *d*-wave superconducting energy gap cannot describe experimental data (please see Fig. 3).

### 3.7. $Ca_{10}(Pt_4As_8)((Fe_{1-x}Pt_x)_2As_2)_5$ single crystal

Ni *et al* [57] discovered two iron arsenide superconductors, $Ca_{10}(Pt_3As_8)(Fe_2As_2)_5$ (the "10-3-8 phase" with $T_c$ = 11 K) and $Ca_{10}(Pt_4As_8)(Fe_2As_2)_5$ (the "10-4-8 phase" with $T_c$ = 26 K). Both phases have been extensively studied [58-61]. As this showed in recent experiments by 10-4-8 phase demonstrates very unusual pairing behaviour [62]. In Fig. 11 we show $B_{c2,c}(T)$ data reported by Mun *et al* [63] for $Ca_{10}(Pt_{4-\delta}As_8)((Fe_{0.97}Pt_{0.03})_2As_2)_5$ ($\delta \approx 0.246$) single crystal. The data fit (with excellent quality) is shown in Fig 10, which also reveals that both deduced ratios, $\frac{2 \cdot \Delta(0)}{k_B \cdot T_c} = 4.9 \pm 0.1$ and $\frac{\Delta C}{C} = 0.89 \pm 0.07$, are within weak-coupling limits of *p*-wave case for this geometry.

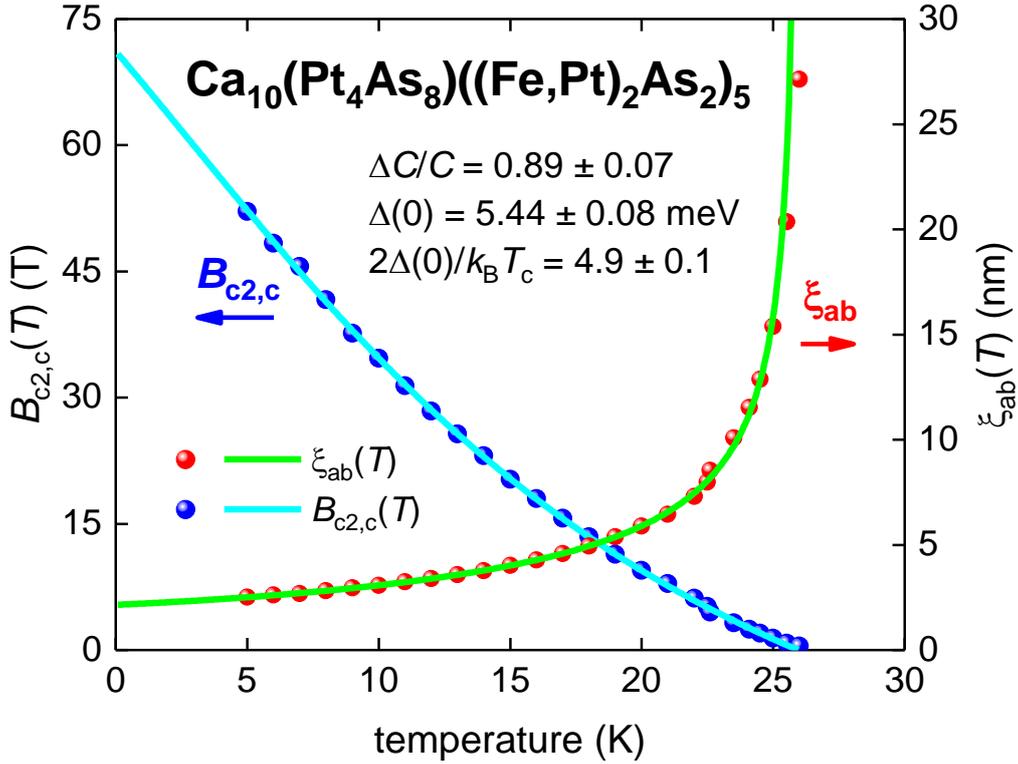

**Figure 11.** *c*-axis upper critical field, $B_{c2,c}(T)$ (raw data reported by Mun *et al* [63]) and data fit to Eqs. 10,11,13-15 for $Ca_{10}(Pt_{4-\delta}As_8)((Fe_{0.97}Pt_{0.03})_2As_2)_5$ ($\delta \approx 0.246$) single crystal. $T_c$ = 26.0 ± 0.3 K and $\xi_{ab}(0)$ = 2.15 ± 0.02 nm. Fit quality is $R$ = 0.9977.

### V. Conclusions

In this paper we describe a general approach to deduce primary parameters of superconductors, i.e. $\xi_{ab}(0)$, $\Delta(0)$, $\Delta C/C$, $T_c$, and $2\Delta(0)/k_B T_c$ by the analysis of experimental upper critical field data, $B_{c2}(T)$, which is defined by strict criterion of $\frac{R(T,B)}{R_{norm}} \to 0$, and which is usually referred as the irreversibility field, $B_{irr}(T)$.

We should stress that $B_{c2}(T)$ data in iron-based superconductors cannot be fitted in the assumption of *s*- or *d*-wave symmetry. We demonstrate that all analysed $B_{c2}(T)$ data in iron-based superconductors can be perfectly fitted to *p*-wave superconducting gap symmetry.

**Acknowledgement**

Author thanks Dr. W. P. Crump (Aalto University) for invaluable help and Dr. J. Hänisch (Karlsruhe Institute of Technology) for providing full raw experimental dataset for Ba(Fe$_{0.992}$Co$_{0.08}$)$_2$As$_2$ thin film.

Author thanks financial support provided by the state assignment of Minobrnauki of Russia (theme "Pressure" No. AAAA-A18-118020190104-3) and by Act 211 Government of the Russian Federation, contract No. 02.A03.21.0006.